\documentclass[preprint,12pt,authoryear]{elsarticle}

\usepackage{pdflscape}
\usepackage{graphicx}
\usepackage{xcolor} 

\usepackage{float} 
\usepackage{graphicx} 
\usepackage{booktabs}  
\usepackage{pifont}    

\usepackage{graphicx}
\usepackage{url}
\usepackage{color, soul}
\usepackage{bm}
\usepackage{amsmath} 
\usepackage{lineno}
\usepackage{pdflscape}
\usepackage{fix-cm}

\usepackage{algorithm}
\usepackage{algpseudocode}
\usepackage{amsmath}

\usepackage{pdflscape}
\usepackage{booktabs}
\usepackage{caption}
\usepackage{graphicx}  

\usepackage{pdflscape} 
\usepackage{float}
\usepackage{multirow}

\usepackage{multirow}
\usepackage{array}
\usepackage{booktabs}
\renewcommand{\arraystretch}{1.4}
\usepackage{tabularx}
\renewcommand{\arraystretch}{1.3}

\usepackage{listings}
\usepackage{xcolor}
\usepackage{caption}
\usepackage{geometry}
\usepackage{makecell}

\usepackage{xcolor}

\definecolor{jsonkey}{rgb}{0.0,0.0,0.6}
\definecolor{jsonstring}{rgb}{0.0,0.5,0.0}
\definecolor{backgroundgray}{rgb}{0.95,0.95,0.95}

\lstdefinelanguage{json}{
    basicstyle=\ttfamily\footnotesize,
    backgroundcolor=\color{backgroundgray},
    morestring=[b]",
    stringstyle=\color{jsonstring},
    literate=
     *{0}{{{\color{black}0}}}{1}
      {1}{{{\color{black}1}}}{1}
      {2}{{{\color{black}2}}}{1}
      {3}{{{\color{black}3}}}{1}
      {4}{{{\color{black}4}}}{1}
      {5}{{{\color{black}5}}}{1}
      {6}{{{\color{black}6}}}{1}
      {7}{{{\color{black}7}}}{1}
      {8}{{{\color{black}8}}}{1}
      {9}{{{\color{black}9}}}{1},
    morekeywords={true,false,null},
    keywordstyle=\color{blue},
}

\usepackage{amssymb}
\usepackage{amsmath}

\journal{International Journal of Applied Earth Observation and Geoinformation}

\begin{document}

\begin{frontmatter}

\title{Context-Aware Visual Prompting: Automating Geospatial Web Dashboards with Large Language Models and Agent Self-Validation for Decision Support}


\author[1]{Haowen Xu\fnref{equal}\textsuperscript{\textdagger,}}
\author[2,3]{Jose Tupayachi\fnref{equal}}
\author[2]{Xiao-Ying Yu\corref{cor1}}

\cortext[cor1]{Corresponding author: yuxiaoying@ornl.gov}
\fntext[equal]{These authors contributed equally to this work.}

\affiliation[1]{organization={Computational Urban Sciences Group, Oak Ridge National Laboratory},
            addressline={1 Bethel Valley Rd}, 
            city={Oak Ridge},
            postcode={37830}, 
            state={TN},
            country={USA}}

\affiliation[2]{organization={Materials Science and Technology Division, Oak Ridge National Laboratory},
            addressline={1 Bethel Valley Rd}, 
            city={Oak Ridge},
            postcode={37830}, 
            state={TN},
            country={USA}}

\affiliation[3]{organization={Industrial and Systems Engineering, The University of Tennessee, Knoxville},
            addressline={851 Neyland Drive}, 
            city={Knoxville},
            postcode={37996}, 
            state={TN},
            country={USA}}

\begin{abstract}
The development of web-based geospatial dashboards for risk analysis and decision support is often challenged by the difficulty in visualization of big, multi-dimensional environmental data, implementation complexity, and limited automation. We introduce a generative AI framework that harnesses Large Language Models (LLMs) to automate the creation of interactive geospatial dashboards from user-defined inputs including UI wireframes, requirements, and data sources. By incorporating a structured knowledge graph, the workflow embeds domain knowledge into the generation process and enable accurate and context-aware code completions. A key component of our approach is the Context-Aware Visual Prompting (CAVP) mechanism, which extracts encodes and interface semantics from visual layouts to guide LLM driven generation of codes. The new framework also integrates a self-validation mechanism that uses an agent-based LLM and Pass@k evaluation alongside semantic metrics to assure output reliability. Dashboard snippets are paired with data visualization codebases and ontological representations, enabling a pipeline that produces scalable React-based completions using the MVVM architectural pattern. Our results demonstrate improved performance over baseline approaches and expanded functionality over third party platforms, while incorporating multi-page, fully functional interfaces. We successfully developed a framework to implement LLMs, demonstrated the pipeline for automated code generation, deployment, and performed chain-of-thought AI agents in self-validation. This integrative approach is guided by structured knowledge and visual prompts, providing an innovative geospatial solution in enhancing risk analysis and decision making.

\end{abstract}

\begin{graphicalabstract}
\begin{figure*}[htb]
 \centering
\includegraphics[width=0.85\textwidth]{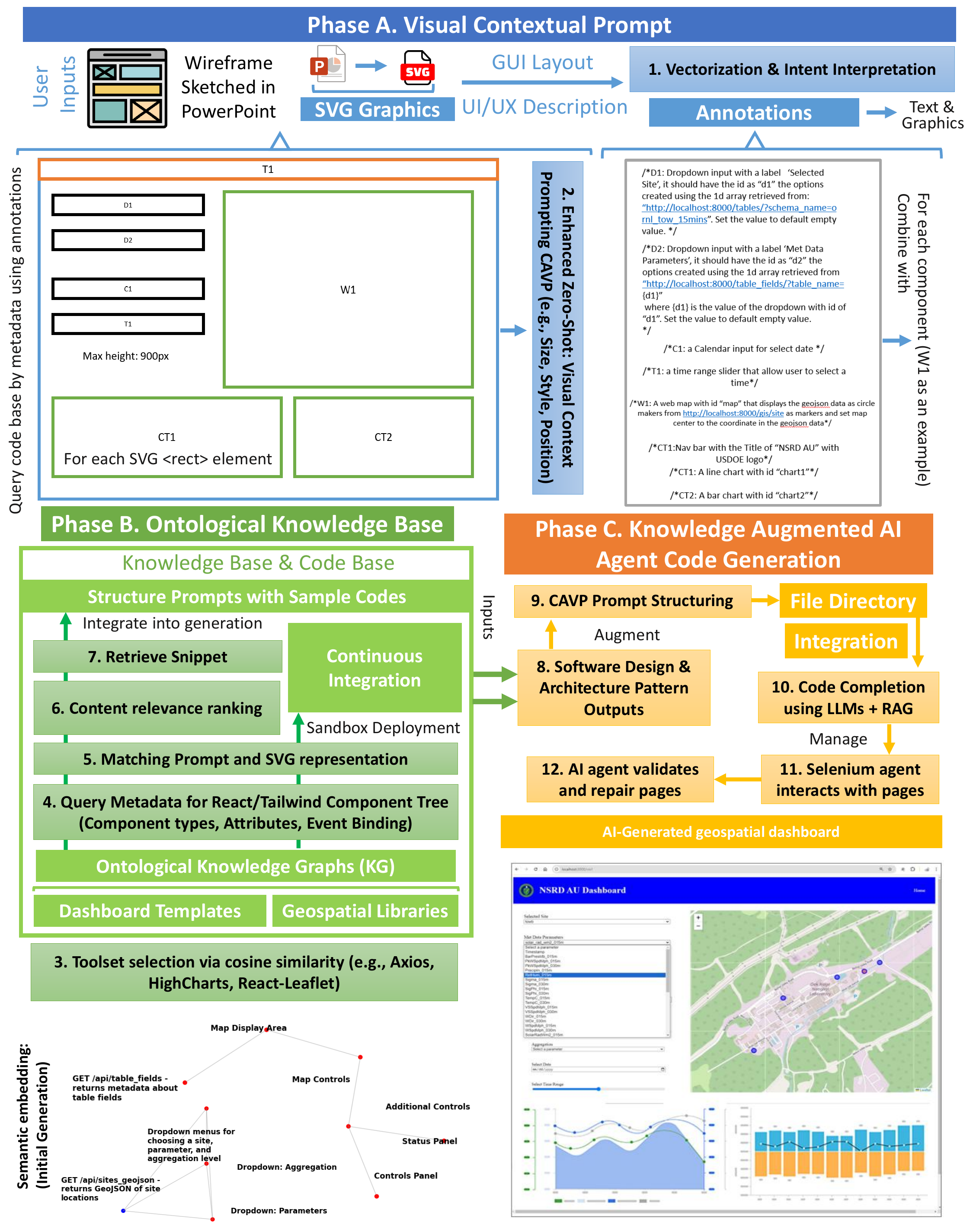}

 \label{fig:concepts}
\end{figure*}
\end{graphicalabstract}

\begin{highlights}

\item Ontological Knowledge Graph and LLMs generate geospatial interactive dashboards.

\item  Context-aware prompting structures the generation of multi-page applications.

\item  Codebase of prompting strategies uses parsed geospatial libraries.

\item  CI/CD implements functionality beyond static-page representations.

\item Automated geospatial dashboard generation leverages CAVP.

\end{highlights}

\begin{keyword}

Risk Analysis and Decision Making,  Large Language Model, Geospatial Dashboards, AI Agent, Semantic Embeddings, Retrieval Augmented Generation (RAG)

\end{keyword}

\end{frontmatter}

\section{Introduction}
\label{Introduction}

Scientific web tools, such as geospatial dashboards, CyberGIS systems, digital twins, and online decision support systems, are essential for researchers and the public to explore, analyze, and use urban and environmental GIS systems with rich environmental data for risk analysis and decision making \cite{ferre2022adoption, dembski2020urban}.
Advances in computing technologies have started the transformation of urban and environmental research by enabling data- and simulation-driven insights for risk analysis and decision support \cite{kadupitige2022enhancing}, fostering collaborative research through data and simulation integration \cite{parashar2019virtual}. The application of geospatial information science is broad, including but not limited to water resource management \cite{souffront2018cyberinfrastructure}, hazard mitigation \cite{mandal2024prime, xu2020web, garg2018cloud}, energy management \cite{jia2019adopting, kim2022design}, emergency response \cite{ li2021emerging, thakur2020covid}, and urban planning and design \cite{alatalo2017two}. Numerous interdisciplinary studies emphasize the transformative impact of artificial intelligence on advancing the capabilities of web applications for interactive environmental research and urban analytics.


Despite progress in scientific web applications, building custom tools, such as cyberGIS platforms to integrate and visualize diverse environmental or urban data, is still complex and resource intensive~\cite{shanjun2024design, siddiqui2024digital, lei2023challenges}. These efforts require expertise in software, design patterns~\cite{fayad2015software}, and data engineering~\cite{kim2017data}, which requires resource-consuming client-server development, database management, and machine learning \cite{ikegwu2022big}. Although a handful of visualization tools (e.g. Kepler.gl) allow users to analyze geospatial data, they have limited functions and require human intervention to build the visualizations. Domain researchers generally lack skills in complex web programming, UI/UX design, and database technologies \cite{li2022bibliometric}, making it difficult for them to implement code development with standards and best practices expected from computational engineering.  Designing, deploying, and maintaining of geospatial apllications is labor intensive, limiting the scalability of platforms ~\cite{shah2024optimizing, mcbreen2002software, liu2015cybergis}. Recent work shows that generative AI (see~\ref{tab:toolsetcode}) can help automate web development for environmental and urban research with static, single output. However, creating complex scientific web applications remains a challenge, because prompting is often inefficient.  Additionally, the code base is generic, trained for non-specialized communications, lack of specificity to serve specific user demands \cite{liang2024can, liukko2024chatgpt}.





\begin{figure*}[htb]
 \centering
\includegraphics[width=\textwidth]{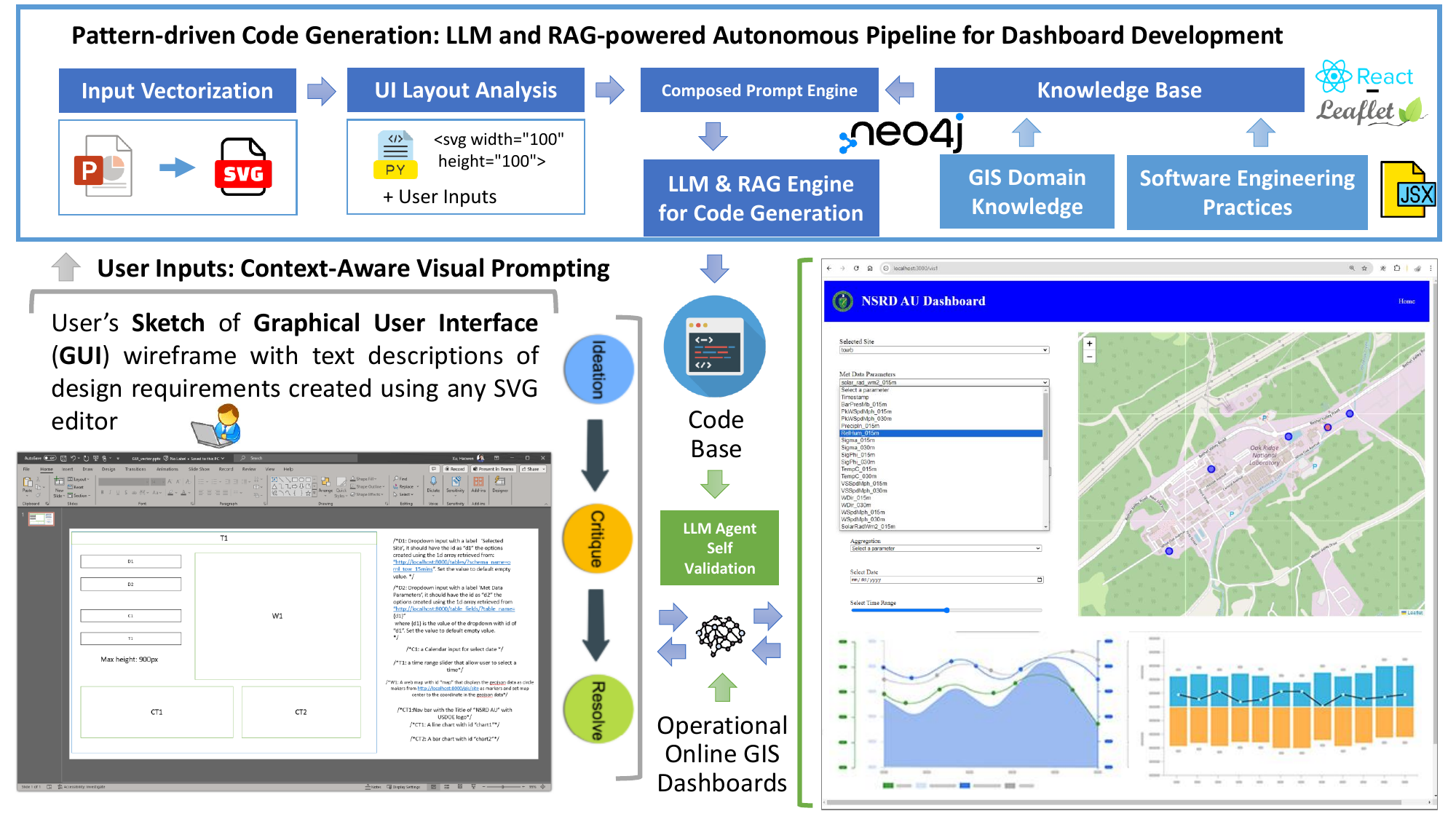}
 \caption{Knowledge-driven framework and annotated wireframes transformation to code for automated platform development and temporal and spatial data visualization.}
 \label{fig:concepts}
\end{figure*}


In this work, we present a novel generative framework that leverages software engineering best practices, domain-specific knowledge, and modern web technologies to automatically generate GIS web applications, including dashboards and analytical tools from user-defined UI wireframes Scalable Vector Graphics (SVGs) with functional requirements expressed through natural language prompts and contextual input depicted in Figure~\ref{fig:concepts}. Using a Python-based, context-aware visual prompting method, our approach interprets wireframes to extract layouts and UI elements, enabling LLMs to generate front-end code guided by Chain-of-Thought (CoT) reasoning, which embeds domain knowledge through an ontological framework and software engineering principles.
A case study is used to show that the framework can automatically build a modular web dashboard for analyzing environmental data.
The dashboard is built upon user input wireframes, utilizing industry-standard frameworks and conventions, like React and MVVM, to assure scalability and maintainability. The pipeline automates setup, builds and serves the app, verifies functionality, and uses AI-assisted repair with validation. This unique framework reduces manual UI design and coding, providing a smart and efficient solution for the creation of geospatial visualization platforms suitable for environmental risk analysis.

\section{Current Implementations \& Literature Review}
\label{Literature Review}
 
LLMs, such as GPT-5 \citep{sun2023gpt}, and DeepSeek \citep{guo2024deepseek}, have transformed software development by enabling advanced code generation. Their proficiency in understanding and generating human language supports reasoning, code synthesis, and problem solving \citep{li2023autonomous}. Ultimately, they provide scalable solutions for automating both repetitive and complex programming needs \citep{meyer2023llm, baldazzi2023fine}. In order to obtain domain-specific applications, such as the creation of UI interfaces in geospatial risk management, further development is needed to adapt LLM models to geospatial data workflows and interactive assessment requirements.


\subsection{A Review of LLM-Based Approaches in Software Development}



Recent studies highlighted the opportunities and challenges of using LLMs in software engineering. \citet{hou2023large} provide an analysis, categorizing LLMs into encoder-only, encoder-decoder, and decoder-only architectures. The growing adoption of decoder-only models for automated code generation and completion has notably reduced manual programming effort. However, challenges remain in handling domain-specific knowledge, improving data quality, and addressing complex software engineering demands. LLMs show promise but still lack the robustness and reliability to fully replace human developers in complex scenarios. For example, HyperAgent \citep{phan2024hyperagent} coodinates agents like the Planner and Code Editor and offers an automated coding workflow. While it demonstrated superior performance in issue resolution and fault localization on benchmarks like SWE-Bench and Defects4J, challenges in scalability and computational overhead remain. Likewise, \citet{xia2024agentless} introduces AGENTLESS, a streamlined approach automating SE tasks through localization and repair. Unlike agent-based systems, AGENTLESS employs a hierarchical structure, reducing integration complexity while achieving 27.33\% on SWE-bench Lite. It automates bug localization and repair efficiently but struggles with cases lacking localization clues and complex reasoning tasks. Although these recent advancements illustrate potential in software engineering for LLM, key challenges in reasoning and scalability remain.



Ongoing research continues to refine LLM-based code generation. \citet{guo2024deepseek} presents DeepSeek-Coder and supports a context window for handling complex tasks, with enhanced cross-file understanding using repository-level data construction and a Fill-in-the-Middle approach. Benchmarks show it outperforms CodeLlama and StarCoder. With a permissive license, \citet{zhang2023planning} proposes Planning-Guided Transformer Decoding (PG-TD), which integrates a planning algorithm with Transformers to improve code generation by leveraging test case results. PG-TD surpasses traditional sampling and beam search, boosting pass rates on competitive programming benchmarks. However, it is computationally intensive and depends on existing test cases, limiting broader applications.

\subsubsection{Evaluating LLMs for Software Development}

LLMs are increasingly evaluated for their capacity in assisting in software engineering procedures. \citet{liang2024can} assessed GPT-4’s ability to replicate automation by constructing full analysis pipelines. While GPT-4 produced good high-level plans, only 30\% of its code ran without changes, showing a lack of domain-specific knowledge. Human oversight remains essential for ensuring accuracy. Likewise, \citet{sandberg2024evaluating} evaluates GPT-4 in full-stack web development, highlighting its efficiency in generating functional applications for simple projects. However, as complexity increases, GPT-4 shows inadequacy in debugging and integration, requiring significant human intervention. \citet{gu2023effectiveness} assesses ChatGPT, CodeLlama, and PolyCoder in domain-specific coding, incurring in API misuse. To address this issue, DomCoder integrates API recommendations with CoT promting to enhance domain-specific automation. However, challenges remain in sourcing domain-specific data and ensuring API consistency. \citet{fan2023large} reviews LLMs, such as GPT, BERT, and Codex, in software engineering, highlighting strengths in code completion, bug detection, and automation. Key challenges include hallucinations and verification issues. It is recognized that future work should focus on better prompt design, integration, and automated testing to improve reliability.




In geospatial applications, \citet{hou2024can} highlight that general-purpose LLMs sometimes fail in code generation or require human guidance. It was found that LLMs could generate hallucinated code, outdated functions, and incompatible dependencies. These challenges become more pronounced when dealing with multi-modal data and platforms such as Earth Engine and Leaflet. Although advanced prompting were used, the identified issues could not be fully addressed. Systematic testing reveals significant shortcomings across commercial and open-source models. Fine-tuning with domain-specific datasets was demonstrated in GEECode-GPT, a Code LLaMA-7B model trained on Google Earth Engine scripts, which significantly improved execution accuracy.

\subsubsection{Front-end App Development from UI Prototypes}
Recent advances in front-end generation and automation focus on bridging UI design and implementation, with new frameworks aiming to generate structured, maintainable code directly from design prototypes. An end-to-end framework~\citet{xiao2024prototype2code} automates front-end code generation from UI design prototypes. Traditional UI-to-code methods often produce fragmented, unstructured code, impacting maintainability and deployment. Prototype2Code detects and corrects UI inconsistencies through linting and constructs a hierarchical layout tree for structured components. Code refinement is performed via UI elements in a Graph Neural Network (GNN)-based classifier before generating modular HTML and CSS code. Benchmarks against CodeFun and GPT-4-based Screenshot-to-Code show superior visual fidelity, readability, and maintainability, which are validated by SSIM, PSNR, and MSE metrics based on ground truth reproduction. They used a case study to confirm reduced manual modifications in dynamic components and improved usability. \citet{manuardi2024images} focused on transforming UI mockups into structured code. Different from text-based coding tools like GitHub Copilot, UI-driven development requires visual processing. They propose an AI system that integrates computer vision and LLMs, using edge detection, contour analysis, and OCR to generate an intermediate representation, which a multi-modal LLM translates into front-end code for Angular and Bootstrap.





\subsection{Limitations and Knowledge Gap}



LLM-based tool-sets for code generation (see Table~\ref{tab:llm_ui_tools} in Appendix) were developed for web development and GIS code generation. They are used within online IDEs, including Claude Designer, Stitch, Cursor, and Base 44, focusing primarily on single-page and static representations. Despite advancements in LLM-powered code generation, drawbacks remain, particularly in scientific and GIS-based web applications. These limitations hinder seamless automation of front-end development, necessitating further research. Such tool-sets depend on prompts, and multimodal Image-to-Code models often fail with cluttered layouts, interactions, and context. Without strong UI understanding, generating structured, modular front-end code remains difficult.
The absence of Software Engineering and lack of integration with industry design patterns (e.g., MVC, MVVM) \citep{xiao2024prototype2code} lead to poor maintainability \citep{ghoshdesign, nguyen2023generative}. LLM-driven workflows rarely incorporate  cotinuos integration deployment (CI/CD) pipelines, software testing, or version control, hindering their practical usability in large-scale development \citep{corona2025question, mendoza2024development}. These weaknesses are further compounded in specialized domains like GIS, where the lack of software engineering practices intersects with domain-specific knowledge gaps. 

Conversational LLMs are unskillful for geospatial and scientific computing due to insufficient training on standards (i.e., GeoJSON, WMS, WFS), web mapping engines, and 3D data visualization \citep{zhang2024bb}. While some fine-tuned models improve geospatial analysis \citep{hadid2024geoscience}, web-based GIS dashboard generation remains largely unexplored. AI-generated code frequently suffers from dependency conflicts and compatibility issues, particularly in GIS and scientific computing \citep{hou2024can, mahmoudi2023development}.These constraints underscore the need for a framework capable of generating dynamic, multi-page interfaces that integrate design logicfram and functional components, enabling automated and scalable generation, integration and continuos deployment (CD). Therefore, Web-based GIS applications require compatibility across libraries (e.g., OpenLayers, Leaflet, Fast API), which LLMs often fail to handle effectively. To bridge these gaps, a robust pipeline is needed to automate the development. Our approach allows users to input mockups using office tool (e.g. Power Point) to export SVGs for code-free development. This work addresses the following identified gaps:

\begin{description}

  \item[Limited UI understanding] Insufficient UI layouts complexity in modular structures, making front-end code generation unreliable.

  \item[Lack of software engineering integration] Current methods do not adopt design patterns, testing, or CI/CD, limiting scalability and maintainability.

  \item[Weak scientific and geospatial support] LLMs lack training on geospatial standards and tools, leading to poor code compatibility and low performance in scientific domain web apps.
\end{description}

\section{Methodology}
\label{Methdology}

To support the retrieval-augmented generation, we construct a structured knowledge base and curate datasets for UI generation using geospatial libraries and React components. The dataset is paired with context-aware visual prompting to verify the method through self-validation mechanisms.


The crafting of high-quality datasets is a foundational step in capturing user requirements and replicating SVG input to accurately generate code completions with structured formatting and robust quality control. Existing approaches to fine-tuning and data set creation often rely on manual testing and human intervention (see Table~\ref{tab:curated_dataset}), which can introduce bias and limit scalability. Although methods like Design2Code~\citep{si2024design2code, gui2025webcode2m} incorporate manual filtering based on selected criteria, these processes remain susceptible to subjective judgment. This highlights the need to develop structured data generation techniques to reduce bias and improve consistency.

Code completion generation can be structured as a pattern-based sequential process, often resembling the \textit{waterfall} model~\citep{royce1970waterfall}, where each stage is completed before the next begins. For example, recent web-UI datasets often begin by collecting raw data from sources such as Common Crawl. These data are then cleaned and filtered by removing noisy or excessively long HTML/CSS files, from which fully loaded screenshots are selected. This approach works well for static or base HTML pages, where the full DOM is available at load time. However, modern frameworks, such as React, Vue, and Angular, use client-side rendering, hiding UI components until runtime. This makes the method unreliable, especially for interactive pages that generate or modify elements dynamically. Our approach begins with the creation of the dataset using a pattern-based templating strategy. This enables creation of multiple page instances with variations in suffixes, prefixes, themes, code completion order, and coding styles. Each entry in the data set contains code completion and file type. We implement unit tests for each page to verify the correctness of components created through the waterfall process and ensure structural and functional integrity. Generated page types include separated pages based on semantic difficulty:







\paragraph{\texttt{Base} (Difficulty: \textbf{1})}
Validates the presence of top-level textual elements, which is expected for validating static or informational base pages.


\paragraph{\texttt{homePage} (Difficulty: \textbf{2})}
Primary textual content, descriptive metadata, banner imagery, and thumbnail elements rendered with labels and navigational links. 


\paragraph{\texttt{Geovisualization} (Difficulty: \textbf{3})}
Evaluates the component's ability to render geospatial data in GeoJSON structures through the openAPI standard. Uses \texttt{React-Leaflet} for map rendering and \texttt{Axios} for HTTP requests. Test coverage includes conditional rendering, empty-state handling, asynchronous loading, or interaction behavior.\\


The difficulty levels reflect the semantic and structural complexity required to pass each test. Level 1 tasks involve minimal logic and serve as basic checks for rendering and static content presence, while higher levels introduce asynchronous operations, external data dependencies, conditional logic, and UI interactions.

\subsection{Context Aware Visual Prompting}

We guide LLMs to generate functional UI code by combining visual layouts with structured domain-specific instructions. At its core, CAVP embeds interface logic and design semantics directly into annotated scalable vector graphic mockups. These vector graphics act as dynamic wireframes, including interpretable guides where each component (e.g., dropdowns, geovisualization, charts) is labeled and spatially organized within the SVG to reflect its role in the application. By pairing these visual prompts with structured API documentation and offering a clear component specification, the LLM is prompted with a full context of both form and function. 


Supported on our codebase, enriched prompts improve layout translations and intentions into styled components with while guided via the knowledge graph. The SVG mock-up input outlines the placement and function of UI elements, such as a site selector, aggregation drop-down, and interactive map panel. The accompanying API schema defines how these elements fetch and process data, which follow the OpenAPI standard. In addition, we include user requirements and file structure, essential to properly handle multi-page imports. This approach provides the selected LLM with comprehensive input, design, functionality, and data behavior, enabling reliable code generation for complex, data-driven dashboards.

\subsection{Ontological Knowledge and Codebase}
\label{subsec:KB_CB}

An ontology is a formal representation of knowledge with defined entities, their properties, and relationships within a domain. In software development, ontologies~\cite{tupayachi2024towards} allow the system to reason about components, workflows, and interactions in a structured, machine-interpretable way. By explicitly modeling concepts, such as UI elements and design patterns. Ontologies offer consistent semantics that supports automated reasoning, inference, and context-aware code generation. We define a structured foundation (T1-T2) used to interpret user design inputs and to guide consistent code generation across tasks. User provided wireframes (visuals, context, and annotations) express software requirements in plain terms. To translate these into structured workflows, an integrated knowledge codebase interprets prompts and supports LLM based code generation. The workflow advances through multiple stages that link conceptual definitions, design interpretation, and code synthesis. The ontology thus serves as the central mechanism ensuring semantic consistency and domain alignment across the entire generation process.

\begin{description}
\item[T1. Content Interpretation]
Users export wireframe files from PowerPoint or Adobe Illustrator in SVG formats, extracting spatial and contextual information to construct HTML layouts and components. Each GUI element such as dropdown menus, charts, and web maps is represented as a vector graphical entity with embedded annotations specifying its function (e.g., data visualization, mapping interface, UI control). These annotations also capture interactions and dependencies, guiding LLMs in event binding for dynamic UI behaviors.

\item[T2. Visual Contextual Info Pairing]
Visual contextual information for each SVG element enables pairing from the reference mockups to the React components. This includes position, size, and style, which are then combined with wireframe annotations to generate structured prompts. By mapping requirements from the mockups to the codebase, not only are individual components aligned, but also layout structures, interdependencies between pages, and event handling logic are captured. This ensures the LLM-generated code adheres to industry standards and preserves the user’s design intent.
\end{description}

\begin{figure*}[htbp]
 \centering
\includegraphics[width=\textwidth]{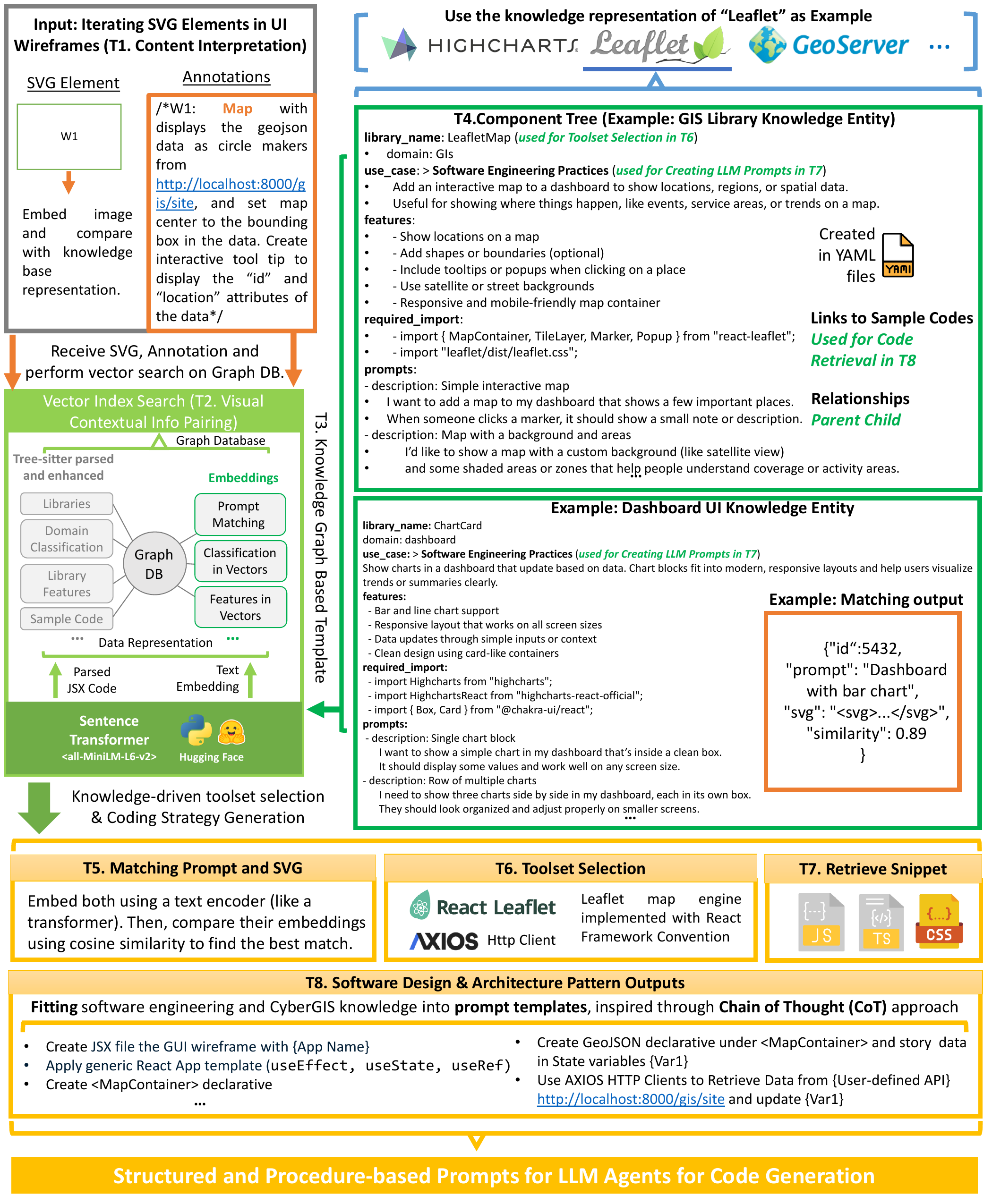}
\caption{Structured knowledge representations for converting plain-language annotations from UI wireframes into structured prompts with technical terminology.}
 \label{fig:KG_method}
\end{figure*}

This codebase leverages knowledge graphs built from UI components (see Appendix~\ref{tab:curated_dataset}), capturing development patterns, imports, and architectures. Graphs are classified via stacks, component strategies, and domain-specific designs, enabling context-aware code generation. The structured knowledge representation (see Figure~\ref{fig:KG_method}) illustrates the graph structure and task-based prompting (T3–T8) used in this workflow.

\begin{description}











\item[T3. Knowledge Graph-Based Component Mapping] This task leverages a knowledge graph to facilitate the mapping of UI components to appropriate web libraries and frameworks. Source code, particularly JSX, is first parsed using Tree sitter and then enriched with metadata including library usage, domain classification, feature descriptions, and representative sample code. This structured information is stored in a graph database, which allows for semantic relationships between libraries, components, and their features to be captured. To support retrieval, the graph is further augmented with vector embeddings representing both the parsed code and textual descriptions. These embeddings enable prompt-based matching, classification in vector space, and retrieval of component features.



\item[T4. Component Tree] Using extracted GUI components and annotations, this task constructs a structured React component tree, preserving semantic correctness, hierarchical relationships, and styles. The wireframe context guides nested structures and component interactions to ensure logical front-end design.

\item[T5. Matching prompt and SVG]
Rather than directly selecting the software stack, this task leverages the pretrained model’s knowledge, input prompting, and retrieved samples to recommend appropriate front-end software stacks. These include the CSS framework, UI libraries, and visualization packages. The templating information is extracted from the underlying knowledge base used during model training (e.g., a Neo4j-backed knowledge graph and post FAISS vector). This approach ensures that the model is aware of the specific completion task.

\item[T6. Toolset Selection]
A post processing filters script, install all the necessary packages, updating, and managing dependencies using appropriate management tools via NPM/NVM. In addition, we mock external dependencies and calls simulating data from APIs or third-party services. This allows components to function and be tested in isolation, ensuring consistent behavior considering the backend dynamics.

\item[T7. Retrieve Snippet]
This task ensures adherence to scalable and modular software design principles by integrating Separation of Concerns (SoC), MVC, and MVVM patterns. It leverages the industry-standard React framework and structured prompt templates to guide LLMs in generating modular code with proper event binding while considering context and file structure.

\item[T8. Software Design \& Architecture Pattern Outputs]
Expanding on T6/T7, this task refines structured prompts for code generation. Sample code from the knowledge base is embedded into prompts, providing procedural instructions to ensure AI generated code adheres to modular design principles for reusability, maintainability, and readability.
\end{description}

\begin{equation}
\label{eq:retrieval}
\text{Retrieve}(P; \mathcal{E}_{\mathcal{K}}) = \left\{ e_i \in \mathcal{K} \;\middle|\; \text{sim}\left(\text{emb}(P), \text{emb}(e_i)\right) \geq \tau \right\}_{i=1}^{k}
\end{equation}

To efficiently retrieve relevant entities at scale, we employ approximate nearest-neighbor search in a compressed domain (see Equation~\ref{eq:retrieval}) using the IVFADC indexing structure~\citep{johnson2019billion}. It enables scalable and efficient retrieval of semantically relevant entities by transforming the user prompt into a dense embedding space, where similarity is computed in a compressed domain to balance accuracy and speed.

\begin{figure*}[htbp]
 \centering
\includegraphics[width=\textwidth]{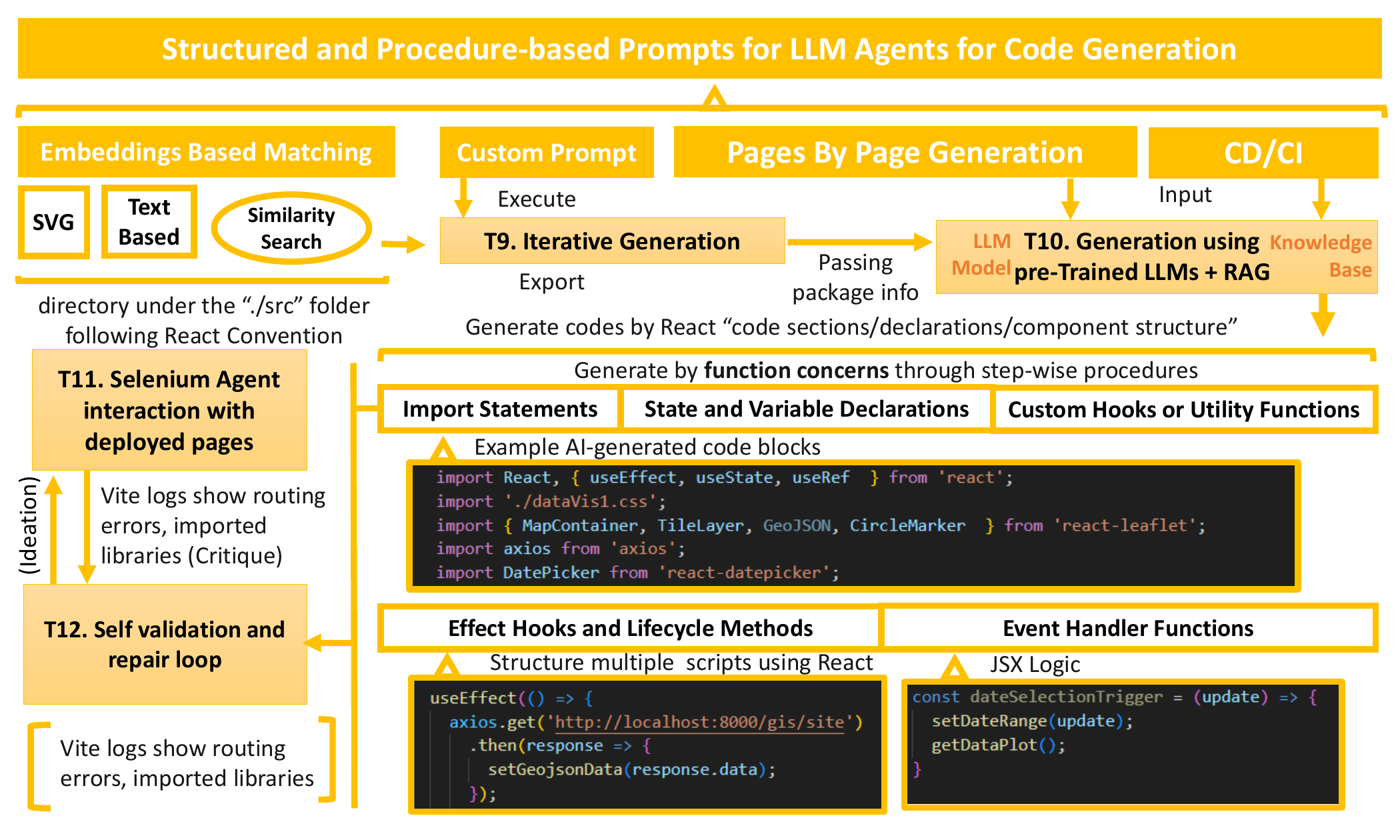}

 \caption{Generation and repair procedure. This process is executed after the ontological knowledge-augmented pipeline produces components aligned with framework best practices.}

 \label{fig:kag}
\end{figure*}

We manage complexity and control output structure by a procedure-based CoT approach with iterative refinement or correction via an Ideation, Decision, and Resolve LLM agent. The execution spans tasks T9 to T12 (see Figure~\ref{fig:kag}).

\begin{description}
    \item[T9. Iterative Generation]

    This task automates the setup and update of dependencies on a per-page basis, leveraging OS-level scripts within a pre-configured Docker environment.  This iterative approach ensures that each page receives focused attention from the model, maintains environment consistency, and simplifies updates while preserving compatibility with required dependencies.

    \item[T10. Generation using pre-trained LLMs + RAG]  
This task uses a procedure based approach for code generation, where Python scripts and the structured codebase guide LLMs to code complete React components in a template driven manner. This ensures adherence to best practices in event handling and state management. The generated code aligns with established architectural and stylistic conventions/

 
\item[T11. Selenium Agent interaction with deployed pages]  
  This task focuses on using Selenium agents to interact with fully deployed React pages, simulating real user behavior to verify UI functionality, routing, and dynamic content rendering. The agents perform automated actions such as clicking buttons, filling forms, and navigating routes, capturing screenshots and DOM states to detect inconsistencies or failures. By interacting with the live application, the system can validate that generated components not only compile but also function correctly in their deployed environment, providing an end-to-end check of usability and correctness. This approach ensures robust testing of the UI, complements static code validation, and provides actionable insights for automated repair or refinement.

\item[T12. Self Validation and repair loop]  
This introduces an iterative pipeline where AI repair agent automatically detect, diagnose, and fix issues in generated React components. By combining Vite-based compilation checks with Selenium-driven functional testing, the system ensures that components are both syntactically correct and behave as intended in the browser. When a component fails validation due to compilation errors, missing imports, or rendering issues, the AI agent proposes corrections, and the code is tested again in the automated loop. This self correcting mechanism significantly improves the reliability of generated code, reduces manual debugging, and ensures that the final application aligns closely with expected UI behavior and functionality.

\end{description}

The pipeline relies on two main processes: 1) an LLM that leverages domain-specific prompt completion pairs for retrieving knowledge snippets through RAG and 2) an agent based LLM for code repair that that further improves evaluated outputs to ensure structural correctness and compilability. By including RAG, the system retrieves knowledge entities along with predefined code snippets (see Figure~\ref{fig:KG_method}), grounding the generation process in semantically relevant examples. This structured context improves accuracy and alignment of the initial front-end code with both user prompts and interface mockups. Additionally, the pipeline orchestrates environment deployment, dependency installation, preprocessing with mocked API calls, and live serving of the application. Once running, a Selenium agent verifies functionality by visiting different pages, while any errors detected during live execution trigger a more detailed verification and agent repair. Repaired code versions are validated with \texttt{esbuild} compilation checks before injection.

\subsection{Repair Agent Using Functional Output}



To enable both scalable generation and dependable refinement of front-end applications, the repair agent (see Figure~\ref{fig:kag}) uses functional shell output analysis to iteratively improve the initial code. While this first generation establishes a structured application layout, it may still require targeted refinement to meet design expectations and compilation needs. 


The AI agent plays a critical role in the reasoning and refinement loop. This agent is responsible for not only producing code candidates but also iteratively critiquing and resolving inconsistencies. Built using LangChain, the agent operates through a structured Ideation loop. At each step, the LLM plans its next action based on current observations, calls the LLM supplying the inconsistencies to receive corrected output. The latter is added into the CI pipline to guide further reasoning based of the Vite/esbuild output. This cycle continues until the system reaches a satisfactory refinement. LangChain manages short and long term memory to orchestrate complex multi-step workflows. Through this, the agent is capable of identifying and generating multiple repair hypotheses based on the outputs against structural and functional criteria. We establish a procedure (see Pseudocode ~\ref{pseudo:agent_eval_vite_dino_math}) for generation and pipeline testing. For each entry in $R$, a prompt is constructed and passed to the LLM $\mathcal{L}$ to completion, which is then deployed and integrated in the application route. An iterative repair loop is then executed for up to $A$ attempts. In each iteration, the development server is started, tests are run, and the set of broken files $B$ is identified. For each broken file and its associated error, the AI agent variant $\mathcal{L}_{\text{agent}}$ attempts up to $F$ fixes, with each proposed fix being validated before saving.

\section{Implementation \& Evaluation}
\label{subsec:impandresults}


We present a curated dataset for testing (see Table~\ref{tab:curated_dataset} in Appendix), featuring diverse instruction-based examples that measure the model’s ability to handle varied instructions prompts. Figure~\ref{fig:kag} depicts the procedural stages of our framework, including prompt parsing, generation, repair, and validation. To assess robustness,

We implement a multi-stage evaluation pipeline guided by   the vite/esbuild output to interpretation, synthesis, and validation of the generated code.


We implement a multi-stage evaluation pipeline guided by the vite/esbuild output to interpretation, synthesis, and validation of the generated code. Several studies have evaluated code generation using various metrics. Traditional ones like BLEU~\cite{papineni2002bleu} are common, but often fail to capture semantic correctness or code-specific structure~\cite{eghbali2022crystalbleu}.To overcome these limitations, CodeBLEU~\cite{ren2020codebleu} incorporated syntax and semantics. In addition, ChrF~\cite{evtikhiev2023out} was used for its ability to reflect nuanced differences in natural and programming languages. For practical validation, we use the \emph{pass@k} metric~\cite{chen2021evaluating}, which estimates the probability that at least one of ``k" completions passes all tests (see Pseudocode~\ref{pseudo:agent_eval_vite_dino_math}).

\subsection{Ablation Study}


To better understand the individual contributions of each pipeline component, we conduct an ablation study that isolates the impact of Codebase RAG on code generation performance (see Table~\ref{tab:pipeline_pass}). The study is performed using different prompting settings, including zero-shot, one-shot, and few-shot, providing a comparative view of performance with and without codebase RAG integration. Among different configurations, the inclusion of RAG leads to measurable improvements in classical code metrics (BLEU, ChrF, TER) and execution-based metrics (pass@k). The experimental setup consists of an Ubuntu instance 24.04.3 LTS with an AMD EPYC 7643 CPU (12 cores, 12 threads), 112 GiB RAM, and 150 GiB storage. GPU computations used an NVIDIA A100 SXM4 80 GB with Driver 570.158.01 and CUDA 12.8. Containers are managed with Docker 28.3.3 and Nvidia Container Toolkit.

We assess execution of LLaMA 3.1 (70B) to systematically analyze the effects of zero-shot, one-shot, and few-shot prompting on code generation throughout different instances of varying degrees of difficulty~\footnote{Execution metrics for other midsize models are also included for comparison. See details at \url{https://huggingface.co/datasets/jtupayac/llm-passk-results}.}. We evaluate the pipeline’s performance at two key stages: initial generation and post-repair compilation after the AI-based repair agent applies the repairs. Success rates are measured using pass@1, pass@5, and pass@10 metrics, providing a clear view of both initial generation quality and improvement after repair.

Our results show that generation compilations achieve moderate pass@1 rates, reflecting challenges of generating fully executable JSX code in complex dashboard pages. The AI repair agent increases success rates in all shots, with improvements observed in pass@1, pass@5, and pass@10. Zero-shot runs benefit substantially from repair, while one-shot and few-shot prompting further enhance the effectiveness of the pipeline, highlighting the interplay between prompt guidance and automated code refinement.


\begin{table}[h!]
\centering
\caption{Summary of Pass@k performance metrics for different variables across prompt strategies. 
Each row corresponds to a specific page type (Geovisualization, Homepage, Base) under a given prompt strategy (One-Shot, Few-Shots, Zero-Shot). 
Columns show the average Pass@k values for k = 1, 3, and 5, indicating the probability that among the top-$k$ generated outputs, at least one is syntactically valid and compiles successfully. 
Best result per page type \emph{and} Pass@k across prompt strategies is in \textbf{bold} with an arrow (↑ = higher is better).}
\label{tab:pipeline_pass}
\begin{tabular}{l l c c c}
\toprule
\textbf{Prompt Strategy} & \textbf{Page Type} & \textbf{Pass@1} & \textbf{Pass@3} & \textbf{Pass@5} \\
\midrule
\multirow{3}{*}{Few-Shots} 
    & Geovisualization & \textbf{0.143 ↑} & \textbf{0.401 ↑} & \textbf{0.620 ↑} \\
    & Homepage          & 0.150            & 0.417            & 0.639            \\
    & Base              & 0.138            & 0.389            & 0.606            \\
\midrule
\multirow{3}{*}{One-Shot} 
    & Geovisualization & 0.139            & 0.392            & 0.609            \\
    & Homepage          & \textbf{0.151 ↑} & \textbf{0.419 ↑} & \textbf{0.642 ↑} \\
    & Base              & \textbf{0.141 ↑} & \textbf{0.395 ↑} & \textbf{0.614 ↑} \\
\midrule
\multirow{3}{*}{Zero-Shot} 
    & Geovisualization & 0.121            & 0.350            & 0.560            \\
    & Homepage          & 0.129            & 0.368            & 0.581            \\
    & Base              & 0.138            & 0.388            & 0.604            \\
\bottomrule
\end{tabular}
\end{table}

\begin{table}[htbp]
\centering
\caption{Aggregated BLEU, ChrF, and TER metrics per page type for each prompt strategy. Best results per page type are in bold with arrows (↑ higher is better for BLEU/ChrF; ↓ lower is better for TER).}
\label{tab:page_type_metrics}
\begin{tabular}{llccc}
\toprule
\textbf{Prompt Strategy} & \textbf{Page Type} & \textbf{BLEU} & \textbf{ChrF} & \textbf{TER} \\
\midrule
\multirow{3}{*}{Few-Shot} 
  & Base      & 25.59 & \textbf{47.83 ↑} & \textbf{59.58 ↓} \\
  & Geovisualization  & \textbf{30.47 ↑} & 61.50 & \textbf{190.51 ↓} \\
  & Homepage  & 56.69 & 70.46 & 69.39 \\
\midrule

\multirow{3}{*}{One-Shot} 
  & Base      & 21.35 & 43.24 & 64.01 \\
  & Geovisualization  & 29.95 & \textbf{61.97 ↑} & 200.21 \\
  & Homepage  & \textbf{83.17 ↑} & \textbf{87.96 ↑} & \textbf{23.20 ↓} \\
\midrule

\multirow{3}{*}{Zero-Shot} 
  & Base      & 18.02 & 41.17 & 67.83 \\
  & Geovisualization  & 21.23 & 53.39 & 232.33 \\
  & Homepage  & 39.74 & 57.65 & 92.95 \\
\bottomrule
\end{tabular}
\end{table}

The Pass@k results (see Table \ref{tab:pipeline_pass}) indicate that prompt strategy strongly affects the likelihood of generating syntactically valid and compilable code. Few-Shot prompting achieves the highest Pass@k values for Geovisualization pages (Pass@1 = 0.143 ↑, Pass@3 = 0.401 ↑, Pass@5 = 0.620 ↑), showing that complex visualizations benefit from multiple examples. One-Shot prompting performs best for Homepage and Base pages (e.g., Homepage Pass@1 = 0.151 ↑, Pass@3 = 0.419 ↑, Pass@5 = 0.642 ↑), suggesting that simpler or more standardized page types can be effectively guided with a single example. Zero-Shot prompting consistently underperforms across all page types, highlighting the need for examples to achieve reliable code generation.

The BLEU, ChrF, and TER metrics provide an additional layer of verification, allowing us to quantitatively assess how closely the generated outputs align with the curated, expert-coded datasets from the waterfall method, and to confirm that the model’s generation follows the expected coding patterns. (see Table \ref{tab:page_type_metrics}) largely confirm these trends. One-Shot prompting produces the highest BLEU and ChrF and the lowest TER for Homepage pages (BLEU = 83.17 ↑, ChrF = 87.96 ↑, TER = 23.20 ↓), aligning with its strong Pass@k performance. Few-Shot prompting achieves the best scores for Geovisualization pages (BLEU = 30.47 ↑, TER = 190.51 ↓), reflecting the benefit of multiple examples for complex outputs. Base pages show smaller differences between strategies, though Few-Shot slightly improves ChrF and TER. Overall, strategies that improve syntactic validity (Pass@k) also tend to enhance similarity to reference code, demonstrating consistency between these complementary evaluation metrics. Since our Few-Shot and One-Shot approaches draw examples from the knowledge graph codebase trhough prompt matching and structured approach, their strong performance in both Pass@k and BLEU/ChrF/TER metrics indicates that the codebase provides high-quality, representative examples. This demonstrates that the codebase effectively captures the coding patterns and structures expected in expert implementations, enabling the model to generalize and produce valid, accurate outputs for new user requests.

\paragraph{Prompt Strategies}

ffective prompting design necessitates multiple refinements to enhance structural consistency and semantic guidance. Key aspects include detailed prompt definitions, precise component descriptions, clear dependency requirements, and well defined layout instructions. The granularity of these details proved central to model performance: more explicit component definitions and structured task instructions consistently provided syntactically correct and functionally coherent React code. These refinements were incorporated iteratively across the Zero-, One-, and Few-Shot settings, ultimately shaping the strategies evaluated in Tables \ref{tab:pipeline_pass} and \ref{tab:page_type_metrics}.


\subsection{Case Study - Meteorological Data Dashboard}

The automatic approach leverages LLM-based prototyping to generate meteorological dashboards, incorporating self-verification technique that focuses exclusively on the generated UI. This verification process supports that the visual elements, such as charts and labels, are displayed correctly and consistently, detecting issues like mislabeling, missing elements, or formatting errors. By automatically validating the accuracy and integrity of the dashboard presentation, the system helps domain experts trust and interact with complex meteorological information, enhancing user-ability to monitor and interpret environmental conditions~\citep{steckler_2025}. The pipeline provides an interactive platform for exploring big environmental data and reaching decisions based potential risks of extreme weather, for instance. The dashboard enables real-time visualization of meteorological variables captured by tower sensors, including temperature, wind speed, humidity, and atmospheric pressure (as shown in Figure \ref{fig:demo-3}). Users can interact with the dashboard to:
\begin{enumerate}
\item Query sensor measurements at different sites, with their locations visualized on the map.
\item Visualize time-series meteorological trends for a large number of parameters over different time periods.
\item Display statistical summaries of temporal data of measured parameters.
\end{enumerate}


\subsection{Baseline Methods}

We compare CAVP against representative baseline approaches that focus on UI code generation from static sources. These methods, such as \textit{ScreenShots2Code}, primarily rely on individual screenshots and file page structures, replicating layouts without leveraging semantic understanding or knowledge graphs. While effective for single-page reconstruction, they lack the mechanisms to capture component relationships or cross-page interactions. The subsequent paragraphs provide a detailed visual comparison, evaluation under Copilot and human-in-the-loop workflows, and an analysis of ontology dependence.



\paragraph{Visual Comparison}

\begin{figure*}[h!]
 \centering
 \includegraphics[height=\textwidth,angle=90]{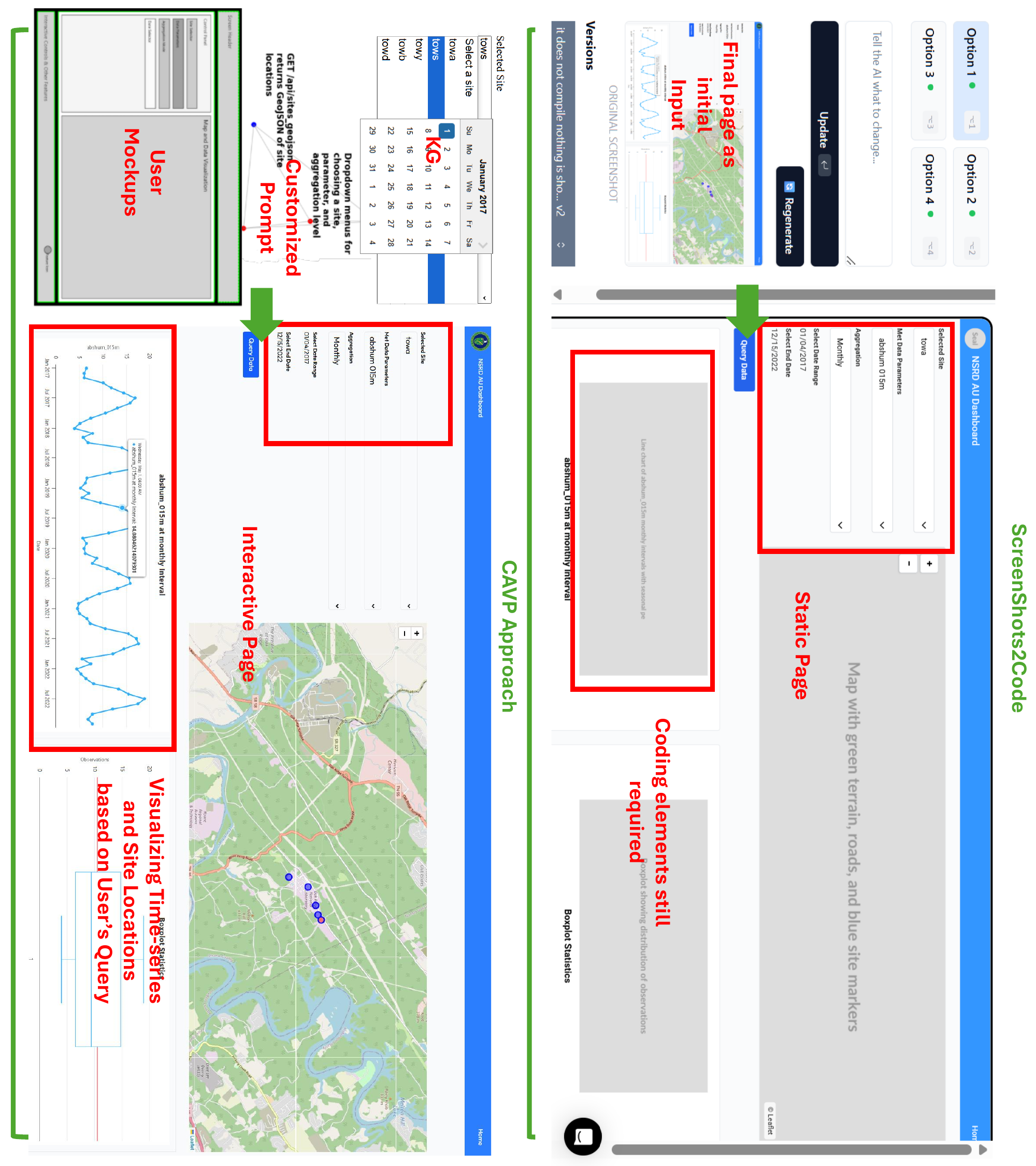}
 \caption{Visual comparison between the proposed CAVP method and the baseline \textit{ScreenShots2Code}. CAVP uses mockups, knowledge graphs, and prompt-based embeddings for semantically-aware UI generation, whereas \textit{ScreenShots2Code} depends on static screenshots and file page structures.}
 \label{fig:demo-3}
\end{figure*}

In this section, we present a visual comparison (see Figure~\ref{fig:demo-3}) between the proposed CAVP method and the baseline \textit{ScreenShots2Code}. The baseline approach relies primarily on static UI screenshots and file page structures, which constrains it to replicating individual static pages without understanding the underlying semantic relationships between UI components. In contrast, CAVP leverages a combination of mockups, knowledge graphs, and prompt-based embeddings to capture both structural and semantic information. This enables the generation of semantically coherent UIs that preserve layout consistency, component hierarchy, and interactivity patterns across multiple pages. As illustrated in Figure~\ref{fig:demo-3}, CAVP is able to produce visually faithful and structurally meaningful UI outputs that go beyond pixel-level replication, highlighting its ability to generalize from the mockup representations and internalize design semantics, unlike the baseline which is limited to surface-level reconstruction.

\paragraph{Copilot Human-in-the-Loop Evaluation} 
We evaluate three primary workflows consisting of Copilot-only, Human-only, and Human-in-the-Loop to measure how model assistance affects developer productivity, code quality, and downstream maintenance cost. Experiments are conducted on a suite of UI engineering tasks drawn from our route and component corpus, comparing model-aided development against additional baselines, including retrieval from a curated snippet corpus, deterministic template generation, and human expert implementations. For each task, we log developer–model interactions, track task completion time, run automated unit and integration tests, and perform static code analysis to capture objective quality metrics. Subjective measures such as developer trust, perceived usefulness, and cognitive load are also collected. Statistical analysis uses mixed-effects models to account for task and participant variability, and qualitative analysis examines failure modes and repair effort.


\paragraph{Ontology Dependence}

The framework’s performance inherently depends on the structure and quality of the underlying ontology, yet this dependence reflects an intentional architectural choice rather than a source of fragility. The ontology operates as a semantic regulator, constraining the language model’s reasoning within domain-consistent boundaries and ensuring interpretability across generated outputs.

While partial misalignments between the ontology and the language model can occur for instance, when relationships are incomplete or concept mappings remain ambiguous, the pipeline is designed to mitigate such cases through retrieval-augmented grounding, task decomposition (T1–T12), and self repair. These stages continuously reintroduce structured cues from the knowledge graph and the existing codebase, thereby reducing semantic drift and preserving internal coherence throughout the generation process. In this sense, ontology dependence is not a limitation but a structural feature that enforces traceability and domain alignment properties that are rarely attainable through unconstrained prompt-based generation alone.


\subsection{Implication for Decision Support}

Critically, the implementation of a codebase ontology underpins the robustness and reliability of this framework. By formally defining the entities, relationships, and constraints inherent to dashboard components, the ontology provides a structured semantic backbone that guides automated code generation. This allows LLMs to reason about dependencies among visual elements, data sources, and interactive workflows.


Ontology-driven code generation still relies heavily on the alignment between LLM reasoning and ontological constraints. While the CoT utilized by the AI Agent and self-verification modules mitigate some risks, gaps between the formal ontology and LLM interpretation can produce syntactically correct but semantically invalid components.The ontological codebase acts as the critical intermediary between abstract user intent (expressed via natural language or mockups) and executable, validated code. Its implementation directly affects the system’s ability to generate maintainable, and semantically consistent dashboards, highlighting its transformative potential.

\paragraph{Generality vs. Domain-specificity}

While our current demonstration focuses on the geospatial domain, the underlying framework is not domain bound. Each experimental setup represents an independent instance that can be adapted to other scientific or engineering contexts. The core pipeline comprising code parsing, example extraction, and visual-context prompting would remains stable across domains.

Only input sources (e.g., domain-specific documentation, visualization libraries, or UI component examples) need to be replaced while quality ensured. Thus, the same workflow can be extended to other applications, upon providing open API exports and mockups targetting those fields.

\begin{table}
\small
\centering
\scriptsize
\caption{LLM-Powered GIS Platforms for Risk and Disaster Management}
\label{tab:llm_gis_risk}
\begin{tabular}{p{3cm}|p{4.5cm}|p{4cm}|p{2cm}}
\hline
 \textbf{LLM Used} & \textbf{GIS Tool or Platform} & \textbf{Risk or Hazard Addressed} & \textbf{Author(s) } \\
\hline
ChatGPT (GPT-4) & Instructor-Agent LLM system integrated with Google Maps for air quality analytics & Wildfire smoke / Air quality & \citet{gao2025instructor} \\
\hline
 Chat bot style-like LLM agent & G.R.O.W. dashboard using NASA FIRMS wildfire data & Wildfires and climate data & \citet{TeamIO_SpaceApps2024} \\
\hline
 Vision-enabled GPT (multi-modal) & AutoS\textsuperscript{2}earch platform for web-based hazard source detection & Industrial hazard (gas leaks) & \citet{zhu2025autos} \\
\hline
 Knowledge-constrained LLM & Flood-aware LLM-GIS platform for risk perception and education & Flood risk via entity constraints and Knowledge Graph relations & \citet{zhu2024flood} \\
\hline
 \textbf{Ontological \& Knowledge guided LLM} & \textbf{CAVP-guided with Knowledge Graph based code synthesis for GIS dashboards} & \textbf{Decision making support in environmental monitoring} & \textbf{This work} \\
\hline
\end{tabular}
\end{table}

The framework supports ongoing environmental monitoring and weather modeling. For leaders and policymakers, it helps improve organizational flexibility. It lets experts like climate scientists, urban planners, and emergency managers build interactive tools without needing advanced programming skills, making geospatial analytics more accessible. Automated dashboard creation with built-in checks enables fast rollout of reliable tools that adapt to changing information and policy needs.

\section{Limitation and Future Work}

We present an original prototype framework that guides LLMs for autonomous and verified code generation in geospatial web applications, focusing on practical deployment in addition to evaluation of LLM performance. Positioned within the larger landscape of geospatial AI-enhanced tools for risk and disaster management (see Table~\ref{tab:llm_gis_risk}). This framework aims to empower risk analysis and decision support in assessing large environmental data through knowledge-guided code synthesis. Unlike many existing systems focused on data analytics or multi-modal hazard detection, our approach emphasizes generating reliable, domain-specific geospatial dashboards that can assist personnel in making timely, informed decisions. However, the framework currently depends on expert-crafted prompt templates. In addition, incorporating refined validation for UI designs will strengthen the alignment between mockup designs and the interface generated. Future work will enhance backend integration, with a focus on data interpretation and servicing.

\section{Conclusion}
\label{Conclusion}

This study presents an ontological knowledge-augmented code generation framework that integrates domain expertise, software engineering principles, and a generative AI repair agent to automate geospatial web application development. Using CAVP, the system transforms user-defined UI wireframes into scalable, maintainable frontend code for complex environmental data visualization. The framework bridges the gap between domain scientists and software engineering best practices, enabling users with minimal web development experience to generate functional geospatial applications. The case study demonstrates its effectiveness in creating multi-page interactive dashboards, including a meteorological data dashboard for visualizing temporal and spatial datasets from long-term records. These AI-generated dashboards support real-time exploration and assessment of large datasets, enhancing usability for scientific research, rish analysis, and ultimately policy-making. To validate the approach, we leverage structured datasets of user requests and implement a modular validation pipeline, including automated testing, code refinement through the AI repair agent, and verification against the original mockups. Overall, our results demonstrate that combining few-shot prompting with an AI repair agent consistently produces the most accurate and reliable code outputs, achieving the highest Pass@1, the best BLEU and ChrF scores, and the lowest TER, highlighting the advantage of including example-based guidance in automated code generation workflows. These components generate applications that are functional to the user-defined designs, providing a scalable, delivering an automated framework for translating UI mockups into validated frontend interfaces, suitable for environmental monitoring, risk analysis, and decision support.

Our work addresses critical technical gaps in current LLM-based frontend generation systems, including insufficient understanding of modular and complex UI layouts, lack of integration with core software engineering principles (e.g., design patterns, testing, and CI/CD), and poor support for scientific and geospatial applications on domain-specific standards and tooling. By introducing the CAVP framework that tightly couples mockup interpretation with code validation and testing, we demonstrate prudent improvements in both structural reliability and design fidelity. Error rates decrease, with TER dropping by 1 to 7 points for the employed LLMs and AI agent. Likewise, visual similarity improves by approximately 8 points, indicating more accurate and maintainable UI code. Larger models show greater gains in both similarity and correctness, while smaller models improve more modestly. However, some metrics begin to plateau, indicating the need for continued tuning and domain adaptation to fully overcome these foundational challenges.

Our framework leverages an ontological knowledge base that formalizes the relationships between UI components, domain-specific standards, and software engineering principles. This ontology serves as a structured intermediary that guides the LLM’s code generation, enabling it to interpret mockups in context, enforce consistency, and reason about dependencies across modules. By encoding both functional and semantic knowledge of UI layouts and geospatial tools, the ontology allows the system to bridge the gap between high-level design intent and executable, validated code.

These advancements are crucial for environmental risk analysis, where accurate, interactive, and reliable geospatial visualization tools are essential for understanding complex multidimensional data and informing timely decision-making. By overcoming key limitations in UI generation, integrating validation, and structuring domain knowledge through ontologies, our framework establishes the kernel for building robust interfaces that can support the dynamic and high-stakes needs of environmental risk assessment and management. Progress in innovative geospatial solutions and applications ultimately advance monitoring, prediction, and mitigation of environmental hazards, fostering more resilient and sustainable communities.









\section*{Data Availability}

The datasets generated and/or analyzed during the current study are available as follows:

\begin{itemize}


\item \textbf{CAVP Dataset}: The curated prompt-completion dataset used for model tuning and validation is accessible at \url{https://huggingface.co/datasets/jtupayac/CAVP_V1}{\texttt{jtupayac/CAVP\_V1}}.

  \item \textbf{Knowledge Base}: The enriched geospatial component and dashboard past projects that use: (JSX GIS Libraries) used for retrieval-augmented generation is accessible at via request.
\end{itemize}


\section*{Declaration of Competing Interest}
The authors declare that they have no known competing financial interests or personal relationships that could have appeared to influence the work reported in this paper. 

\section*{CRediT Authorship Contribution Statement}
\begin{itemize}
  \item \textbf{Haowen Xu:} Conceptualization, Methodology, Writing original draft.
  \item \textbf{Jose Tupayachi:} Methodology, Fine tuning, Data curation, Software development, AI agents, Writing original draft, Review \& editing.
  \item \textbf{Xiao-Ying Yu:} Conceptualization, Writing original draft, Supervision, Project administration, Funding acquisition.
\end{itemize}

\section*{Acknowledgments}


\begin{description}

    \item \noindent This manuscript has been authored by UT-Battelle, LLC,
under contract DE-AC05-00OR22725 with the US DOE.
The US government retains and the publisher, by accepting
the article for publication, acknowledges that the US
government retains a nonexclusive, paid-up, irrevocable,
worldwide license to publish or reproduce the published
form of this manuscript, or allow others to do so, for US
government purposes. DOE will provide public access to
these results of federally sponsored research in accordance
with the DOE Public Access Plan
(https://www.energy.gov/doe-public-access-plan). 
\item \noindent This research used resources from the ORNL Research Cloud Infrastructure at the Oak Ridge National Laboratory, which is supported by the Office of Science of the U.S. Department of Energy under Contract No. DE-AC05-00OR22725. 

\item \noindent\textsuperscript{\textdagger}Current address: School of Built Environment, UNSW Sydney, Red Centre (H13) West, University Mall, Kensington, NSW 2033, Australia

\end{description}


\bibliographystyle{plainnat}
 \bibliography{bib_file_elsevier}

\appendix


\newpage

\section{Toolset for Code Generation}\label{tab:toolsetcode}

\begin{table}[H]
\centering
\scriptsize
\caption{Overview of LLM-Based Toolsets for web development and potential GIS code generation}
\label{tab:llm_ui_tools}
\begin{tabular}{p{2.2cm}|p{2.4cm}|p{5.8cm}|p{2.4cm}}
\hline
\textbf{Tool} & \textbf{Focus} & \textbf{Key Features} & \textbf{Reference} \\
\hline
GPT-5 / ChatGPT & General UI generation & Text-to-code generation; useful for prototyping but error-prone and often hallucinates. & \citet{meyer2023llm} \\
\hline
GitHub Copilot & Code completion & Boosts dev productivity; up to 40\% of suggestions shown to be insecure. & \citet{pearce2022copilot} \\
\hline
Prototype2Code & HTML/CSS from mockups & Converts sketches to static HTML using LLM + layout heuristics; lacks interactivity. & \citet{xiao2024prototype2code} \\
\hline
UICoder & SwiftUI from text & Uses compiler and vision feedback; produces mostly simple, static apps. & \citet{wu2024uicoder} \\
\hline
GeoCode-GPT & GIS code generation & Trained on Earth Engine; improves performance on GIS scripts. & \citet{hou2024geocodegpt} \\
\hline
Claude& Dialog/code reasoning & Strong at multi-step reasoning; weaker at long-form UI code, controllable. & \citet{anthropic2023claude} \\
\hline
Gemini & Web-integrated assistant & Excels at web-connected tasks; generates React code able to fine tune based on Open-source Gemma. & \citet{google_gemini2025cli} \\
\hline
DeepSiteV3 & Code generation LLM & Open-source code-specialized model; supports frontend stacks (e.g., Vue, React). Early results promising. & \citet{guo2024deepseek} \\
\hline

\textbf{CAVP} & \textbf{Code generation LLM} & \textbf{Fine-tuned using a structured knowledge base built on open-source, code-specialized models with CD/CI.} & \textbf{This Work} \\
\hline

\end{tabular}
\end{table}

\section{Dataset Overview}

\begin{table}[H]
\centering
\scriptsize
\renewcommand{\arraystretch}{1.3}
\caption{Overview of UI Generation Datasets}
\label{tab:curated_dataset}
\begin{tabular}{p{2cm}|p{2.5cm}|p{2.0cm}|p{5.9cm}|p{2.0cm}}
\hline
\textbf{Tool Name} & \textbf{Dataset} & \textbf{Testing} & \textbf{Notes} & \textbf{Author} \\
\hline

\textbf{WebCode2M} &
2,563,905 pages (HTML/CSS) &
768 pages (3×256 short/mid/long) &
Real web pages (Common Crawl); includes layout and DOM labels. &
\citet{gui2025webcode2m} \\
\hline

\textbf{Vision2UI} &
16,000 pages &
2,000 pages &
Real pages (Common Crawl); includes element bounding boxes and rendered screenshots. &
\citet{gui2024vision2ui} \\
\hline

\textbf{Design2Code} &
7,000 manually filtered pages &
484 pages &
Filtered for safe content and valid formatting; high-quality real-world HTML/CSS examples. &
\citet{si2024design2code} \\
\hline

\textbf{ZSPrompt} &
50 manually curated prompts &
50 generated UIs &
UIs generated via zero-shot GPT-4 and rated by 3 Prolific workers for functionality, aesthetics, and errors. &
\citet{kolthoff2024zero} \\
\hline

\textbf{WebGen-Bench} &
6,667 NL instructions &
101 instructions &
Instruction-to-UI task; test set includes human-written target UIs and functional test cases. & \citet{lu2025webgen}
 \\
\hline

\textbf{WebSight} &
823,000 pages (synthetic) &
484 pages &
Synthetic LLM-generated UI, webpages curated for test. Includes code + rendered images. & \citet{laurenccon2024unlocking}
 \\
\hline

\textbf{CAVP} &
\textbf{1141 Searchable components, 5 GIS libraries (Code Base)} &
\textbf{200 Independent pages}  &
\textbf{Synthetic LLM-generated dashboard and curated data set for test. Includes code + SVG mockups.} & This Approach
\\
\hline
\end{tabular}
\end{table}

\section{Pipeline Pseudocode}

\begin{algorithm}[H]
\small
    \caption{Iterative generation: concatId groups test pages that originate from the same instance.}
\label{pseudo:agent_eval_vite_dino_math}
\begin{algorithmic}[1]
\Require Dataset $D$, target ID $c$, LLM model $\mathcal{L}$, max attempts $A$, max fixes $F$

\State $R \gets \{e \in D \mid e.\texttt{concatId} = c \}$

\ForAll{$e \in R$} 
    \State $p \gets \text{BuildPrompt}(e)$
    \State $code \gets \mathcal{L}(p)$
    \State \text{Deploy}$(code)$
    \State \text{Mock fetch-apis}$(code)$
\EndFor

\For{$a = 1$ to $A$}
    \State \text{StartServer}()
    \State \text{RunTests}(\texttt{routes})
    \State $B \gets \text{BrokenFiles : Repair}()$
    \If{$B = \emptyset$} \textbf{break} \EndIf
    \ForAll{$(f, err) \in B$}
        \For{$t=1$ to $F$}
            \State $fix \gets \mathcal{L}_\text{agent}(f, err)$
            \If{$\text{Valid}(fix)$}
                \State \text{Save}$(fix)$
                \State \textbf{break}
            \EndIf
        \EndFor
    \EndFor
\EndFor
\end{algorithmic}
\end{algorithm}

\section{IVFADC}

Each embedding vector \( y \in \mathbb{R}^d \) from the knowledge base is quantized using a two-level quantizer.

\[
q(y) = q_1(y) + q_2(y - q_1(y))
\]
where \( q_1: \mathbb{R}^d \rightarrow \mathcal{C}_1 \) is a coarse quantizer and \( q_2: \mathbb{R}^d \rightarrow \mathcal{C}_2 \) refines the residual. Given a query \( x = \text{emb}(P) \), the asymmetric distance computation retrieves approximate neighbors by minimizing:
\[
L_{\text{ADC}} = \underset{i}{\text{k-argmin}} \;\left\| x - q(y_i) \right\|^2
\]

Inverted file indexing restricts the search to the nearest \( \tau \) coarse centroids from \( \mathcal{C}_1 \), forming a shortlist:
\[
L_{\text{IVF}} = \underset{c \in \mathcal{C}_1}{\tau\text{-argmin}} \;\left\| x - c \right\|^2
\]

The final shortlisted candidates are as follows:
\[
L_{\text{IVFADC}} = \underset{i:\, q_1(y_i) \in L_{\text{IVF}}}{\text{k-argmin}} \;\left\| x - q(y_i) \right\|^2
\]

\end{document}